\documentclass[runningheads,a4paper]{llncs}

\usepackage{amsfonts,amssymb,amsmath}
\usepackage{natbib}

\usepackage{wrapfig}
\usepackage{llncsdoc}
\usepackage{epsfig,psfrag}
\usepackage{latexsym}
\usepackage{longtable}
\usepackage{tikz,pgf}
\usepackage{subfig}
\usepackage{wrapfig}

\usepackage{graphicx}
\usepackage{epstopdf}
\usepackage{color}

\newcommand{\bqn}{\begin{eqnarray}}
\newcommand{\eqn}{\end{eqnarray}}
\newcommand{\bq}{\begin{eqnarray*}}
\newcommand{\eq}{\end{eqnarray*}}
\usepackage[ruled,vlined]{algorithm2e}

\usepackage{url}
\usepackage{hyperref}
\hypersetup{colorlinks,%
citecolor=black,%
filecolor=blue,%
linkcolor=red,%
urlcolor=blue,%
pdftex}

\newcommand{\blue}[1]{{\color{blue} #1}}

\begin{document}

\title{Sparse Network Modeling}
 \titlerunning{Logistic Regression}
 
\author{Moo K. Chung
 }
\institute{
University of Wisconsin-Madison, USA\\
\vspace{0.3cm}
\blue{\tt mkchung@wisc.edu}
}
\authorrunning{Chung}

\maketitle

\begin{center}
July 29, 2020
\end{center}

\pagenumbering{arabic}

\index{sparse!network}

There have been many  attempts to identify high-dimensional network features via multivariate approaches \citep{chung.2013.MICCAI,lerch.2006,he.2007,worsley.2005.neural,he.2008}. Specifically, when the number of voxels or nodes, denoted as $p$, are substantially larger than the number of images, denoted as $n$, it produces an under-determined model with infinitely many possible solutions. The {\em small-$n$ large-$p$ problem} is often remedied by regularizing  the under-determined system with additional sparse penalties. 

Popular sparse network models include sparse correlations \citep{lee.2011.TMI,chung.2013.MICCAI, chung.2015.TMI, chung.2017.IPMI}, LASSO \citep{bickel.2008,peng.2009,huang.2009,chung.2013.MICCAI},
sparse canonical correlations \citep{avants.2010} and graphical-LASSO \citep{banerjee.2006,banerjee.2008,friedman.2008,huang.2009,huang.2010,mazumder.2012,witten.2011}. These popular sparse models require optimizing $L1$-norm penalties, which has been the major computational bottleneck for solving large-scale problems. Thus, many existing sparse brain network models in brain imaging have been restricted to a few hundreds nodes or less. 2527 MRI features used in {a LASSO model} for Alzheimer's disease \citep{xin.2015} is probably the largest number of features used in any sparse model in the brain imaging literature.

\section{Why sparse network models?}
\index{sparse models}

If we are interested quantifying the measurements in every voxel in an image simultaneously, the standard procedure is to set up a multivariate general linear model (MGLM), which generalizes widely used univariate GLM  by incorporating vector valued responses and explanatory variables \citep{anderson.1984, friston.1995.MGLM, worsley.1996, worsley.2004, taylor.2008, chung.2010.ni}. Hotelling's $T^2$-statistic is a special case of MGLM and has been mainly used for inference on surface shapes and deformations \citep{thompson.1997, joshi.1998, cao.1999.stat, gaser.1999, chung.2001.ni}.

Let ${\bf J}_{n \times p} = (J_{ij})$ be the measurement matrix, $J_{ij}$ is the measurement for subject $i$ at voxel position $j$. The subscripts denote the dimension of matrix.  We can think $J_{ij}$ as either Jacobian determinant, fractional anisotropy values or fMRI activation. Assume there are total $n$ subjects and $p$ voxels of interest.  The measurement vector at the $j$-th voxel is denoted as ${\bf x}_j=(J_{1j}, \cdots, J_{nj})^{\top}$. The measurement vector for the $i$-th subject is denoted as ${\bf y}_i = (J_{i1}, \cdots, J_{ip})$, which is expected to be distributed identically and independently over subjects. Note that 
$${\bf J} = ({\bf x}_1, \cdots, {\bf x}_p) = ({\bf y}_1^{\top}, \cdots, {\bf y}_n^{\top})^{\top}.$$ 
We may assume the  covariance matrix of ${\bf y}_i$ to be
$$\mathbb{V}({\bf y}_1) = \cdots = \mathbb{V}({\bf y}_n) = {\bf \Sigma}_{p \times p} = (\sigma_{kl}).$$
With these notations, we set up the following MGLM over all subjects and across different voxel positions:
\bqn {\bf J}_{n \times p} = {\bf X}_{n \times k}{\bf B}_{k \times p} + {\bf Z}_{n \times q}{\bf G}_{q \times p} + {\bf U}_{n \times p}{\bf \Sigma}_{p \times p}^{1/2},\label{eq:multivariate}\eqn
where $\bf X$ is the matrix of contrasted explanatory variables while $\bf B$ is the matrix of unknown coefficients to be estimated. Nuisance covariates of non-interest are in the matrix $\bf Z$ and the corresponding coefficients are in the matrix $\bf G$. The components of Gaussian random matrix $\bf U$ are independently distributed with zero mean and unit variance. The symmetric matrix $\bf \Sigma^{1/2}$ is the square-root of the covariance matrix accounting for the spatial dependency across different voxels. In MGLM (\ref{eq:multivariate}), we are  interested in testing the null hypothesis 
$$H_0: {\bf B} = 0.$$ 
The parameter matrices in the model are estimated via the least squares method. The resulting multivariate test statistics are called the Lawley-Hotelling trace or Roy's maximum root. When there is only one voxel, i.e. $p=1$, 
these multivariate test statistics collapses to  Hotelling's $T^2$-statistic \citep{worsley.2004}.

\begin{figure}[t]
\centering
\includegraphics[width=0.7\linewidth]{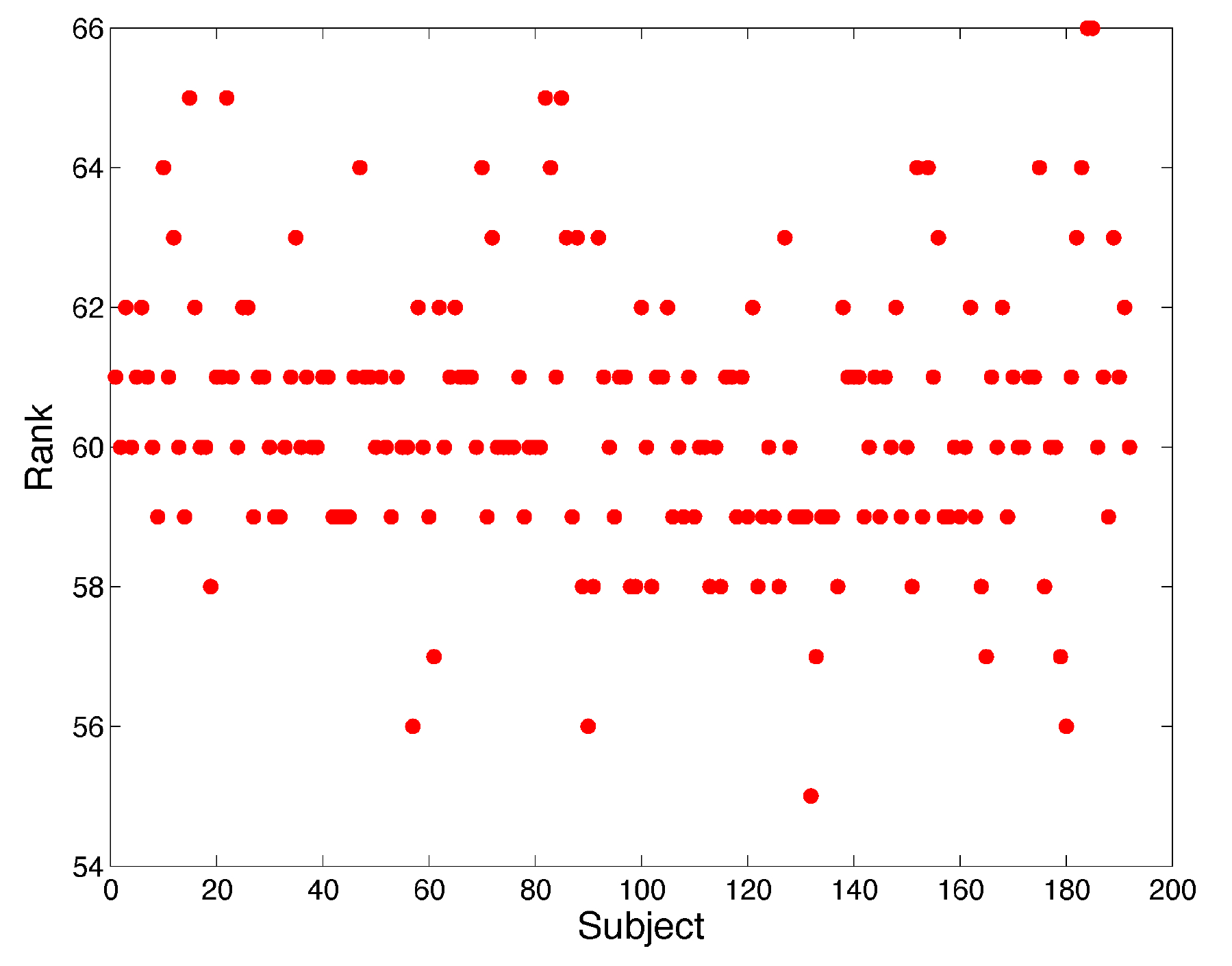}
\caption{The rank of 80-nodes fMRI correlation matrices for 192 subjects published in \citep{qiu.2015}. None of correlation matrix is of full rank and not invertible. Many brain regions show pairwise correlations.}
\label{fig:rank}
\end{figure}

Note that MGLM (\ref{eq:multivariate}) is equivalent to the assumption that ${\bf y}_i$  follows multivariate normal with some mean $\mu$ and covariance ${\bf \Sigma}$, i.e., ${\bf y}_i \sim N(\mu, {\bf \Sigma})$. Then neglecting constant terms, the log-likelihood function $L$  of ${\bf y}_i$ is given by
\bq L  (\mu, {\bf \Sigma}) &=&  \log \det {\bf \Sigma}^{-1} - \frac{1}{n}\sum_{i=1}^n ( {\bf y}_i - \mu)^{\top} {\bf \Sigma}^{-1} ({\bf y}_i - \mu). \label{eq:log1}\eq
By maximizing the log-likelihood, MLE of
$\mu$ and $\bf \Sigma$ are given by 
\bqn \widehat{\mu} &=& \bar{\bf y}_i = \frac{1}{n}\sum_{i=1}^n {\bf y}_i  \nonumber \\
 \widehat{\bf \Sigma}&=& \frac{1}{n}\sum_{i=}^n  ({\bf y}_i - \bar{\bf y}_i)^{\top} ({\bf y}_i - \bar{\bf y}_i).\label{eq:sigmaMLS2} \eqn
For a notational convenience, we can center the measurement  ${\bf y}_i$ such that
$${ \bf y}_i \leftarrow {\bf y}_i  -    \bar{\bf y}_i.$$
We are basically centering the measurements by subtracting the group mean over subjects. Then MLE (\ref{eq:sigmaMLS2}) can be written in a more compact form 
\bqn \widehat{\bf \Sigma} = \frac{1}{n} {\bf J}^{\top}_{p \times n} {\bf J}_{n \times p}. \label{eq:sigmaMLS2}\eqn

However, there is a serious defect with  MGLM (\ref{eq:multivariate}) and its MLE (\ref{eq:sigmaMLS2}); namely the estimated covariance matrix $\widehat{\bf \Sigma}$ is positive definite only for $n \geq p$ \citep{friston.1995.MGLM, schafer.2005}. 
${\bf J}^{\top} {\bf J}$ becomes rank deficient for $n < p$. In most imaging studies, there are more voxels than the number of subjects, i.e., $n < p$. Even when $n > p$, for various reasons, correlation and covariance matrices may not be full rank (Figure \ref{fig:rank}). When $\widehat{\bf \Sigma}$ is singular, we do not  properly have the inverse of $\widehat{\bf \Sigma}$, which is the precision matrix often needed in partial correlation based network analyses \citep{lee.2011.TMI}. This is the main reason MGLM was rarely employed over the whole brain region and researchers are still using mostly univariate approaches in imaging studies.

\subsection{Why sparse network?}
\index{sparse networks}

The majority of functional and structural connectivity studies in brain imaging are usually  performed following the standard analysis framework \citep{gong.2009, hagmann.2007, fornito.2010, zalesky.2010}. From 3D whole brain images, $n$ regions of interest (ROI) are identified and serve as the nodes of the brain network. Measurements at ROIs are then correlated in a pair-wise fashion to produce the connectivity matrix of size $n \times n$. The connectivity matrix is then thresholded to produce the adjacency matrix consisting of zeros and ones that define the link between two nodes. The binarized adjacency matrix is then used to construct the brain network. Then various graph complexity measures such as degree, clustering coefficients, entropy, path length, hub centrality and modularity are defined on the graph and the subsequent statistical inference is performed on these complexity measures.

For  a large number of nodes, simple thresholding of correlation will produce a large number of edges which makes the interpretation difficult. For example,  for $3 \times 10^5$ voxels in an image, we can possibly have  a total  of  $9 \times 10^{10}$ directed edges in the graph. 
For this reason we used the sparse data recovery framework in obtaining  a far smaller number of significant edges.

\section{Sparse likelihood}
\index{graphical-LASSO}
\index{sparse likelihood}
\index{likelihood!sparse}
\index{likelihood}

 \begin{figure}[t!]
\centering
\includegraphics[width=0.8\linewidth]{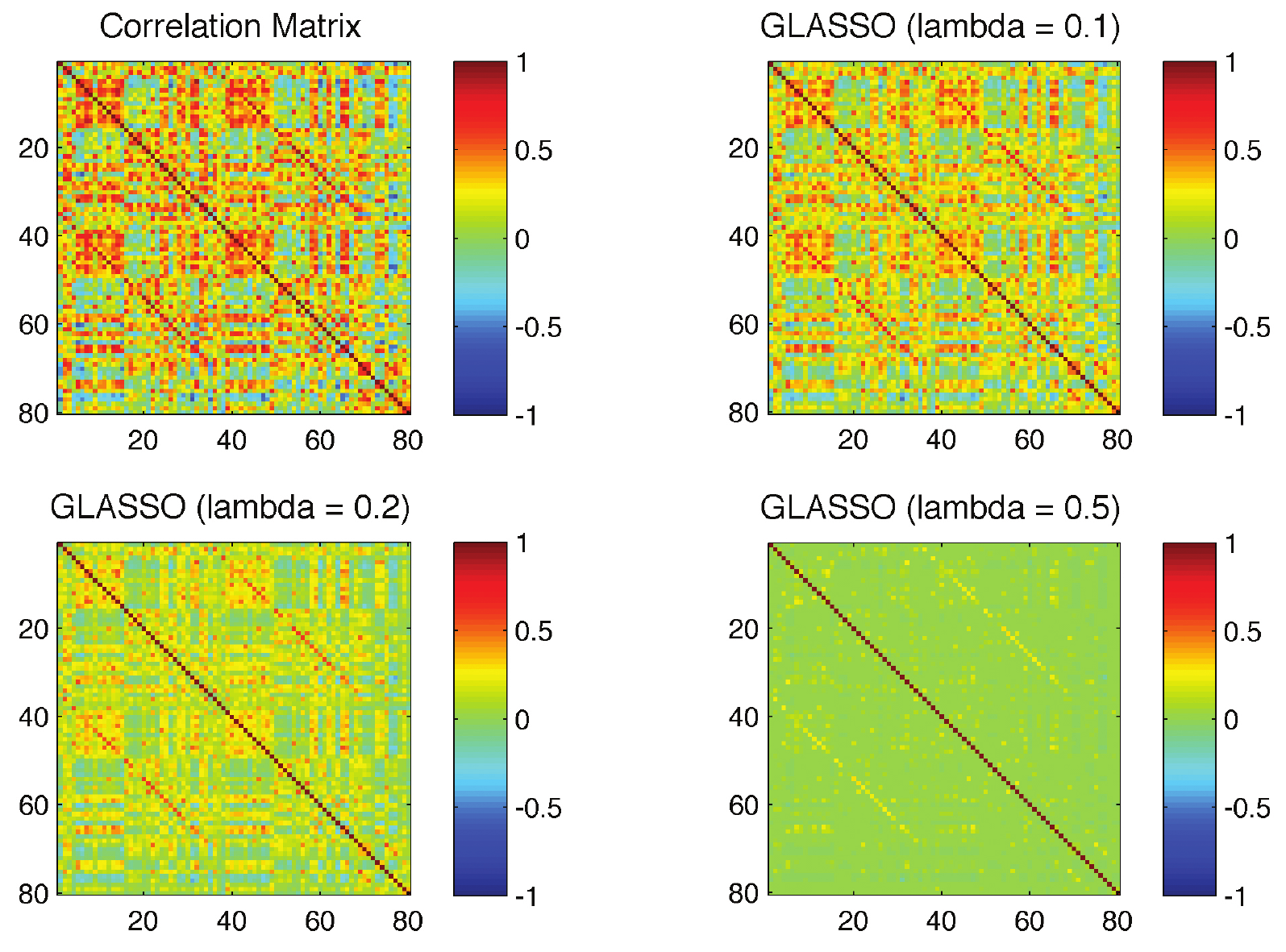}
\caption{Graphical-LASSO estimation on an original singular correlation matrix from study \citet{qiu.2015}. The original correlation matrix has the rank of 61 indicating approximately 19 nodes out of 80 nodes are dependent of other nodes. As the sparse parameter $\lambda$ increases, we see more sparsity and diagonal dominance that makes the estimated sparse correlation matrix to be more positive definite.}
\label{fig:GLASSO}
\end{figure}

Beyond sparse regression, others have proposed the likelihood methods. To remedy the small$-n$ and large-$p$ problem, the likelihood  is  regularized with a L1-norm penalty. If we center the measurements ${\bf y}_i$, the log-likelihood can be written as 
\bq L({\bf \Sigma}) &=& \log \det {\bf \Sigma}^{-1} - \frac{1}{n}\sum_{i=1}^n  {\bf y}_i^{\top} {\bf \Sigma}^{-1} {\bf y}_i \\
&=& \log \det {\bf \Sigma}^{-1} - \mbox{ tr} \big ( {\bf \Sigma}^{-1} S \big),\eq
 where $S = \frac{1}{n}\sum_{i=1}^n  {\bf y}_i^{\top} {\bf y}_i$ is the sample covariance matrix. We used the fact that the trace of a scalar value is equivalent to the scalar value itself and $\mbox{ tr} (AB) = \mbox{ tr} (BA)$ for matrices $A$ and $B$.

To avoid the small-$n$ large-$p$ problem, we penalize the log-likelihood with L1-norm penalty: 
\bqn L({\bf \Sigma}^{-1}) = \log \det {\bf \Sigma}^{-1} -  \mbox{ tr} \Big(  {\bf \Sigma}^{-1} S\Big) - \lambda \| {\bf \Sigma}^{-1} \|_1, \label{eq:banerjee2}\eqn
where $\| \cdot \|_1$ is the sum of the absolute values of the elements. We made the likelihood as a function of ${\bf \Sigma}^{-1}$ to simply emphasize that we are trying to estimate the inverse covariance matrix. The penalized log-likelihood is maximized over the space of all possible symmetric positive definite matrices. (\ref{eq:banerjee2}) is a convex problem and it is usually solved using the graphical-LASSO (GLASSO) algorithm \citep{banerjee.2006,banerjee.2008,friedman.2008,
huang.2010,mazumder.2012}. The tuning parameter $\lambda > 0$ controls the sparsity of the off-diagonal elements of the inverse covariance matrix. By increasing $\lambda>0$, the estimated inverse covariance matrix becomes more sparse (Figure \ref{fig:GLASSO}).

GLASSO is a fairly time consuming algorithm \citep{friedman.2008,huang.2010}. Solving GLASSO for 548 nodes, for instance, may take up to 6 minutes on slow desktop computers if fast algorithms like \citet{hsieh.2013} is not used. If ${\bf \Sigma}^{-1}_i(\lambda)$ is the estimated inverse sparse covariance for group $i$ at given sparse parameter $\lambda$, we are interested in testing the equivalence of inverse covariance matrices between the two groups at fixed $\lambda$, i.e., 
$$H_0: {\bf \Sigma}_1^{-1} (\lambda)= {\bf \Sigma}_2^{-1}(\lambda).$$

\subsection{Filtration in graphical-LASSO}
\index{sparse likelihood}
\index{likelihood!sparse likelihood}

The solution to graphical-LASSO has a peculiar nested topological structure. Let $\Sigma^{-1}(\lambda) = (\sigma^{ij}(\lambda))$ be the inverse covariance estimated from graphical-LASSO.  Let $A(\lambda) = (a_{ij})$ be the corresponding adjacency matrix given by 
\bqn a_{ij}(\lambda) = 
\begin{cases}
1 &\; \mbox{  if  } \widehat{\sigma}^{ij} \neq 0;\\
0 & \; \mbox{ otherwise.}
\end{cases} \label{eq:Aadj}
\eqn
The adjacency matrix $A$ induces a graph $\mathcal{G}(\lambda)$ consisting of $\kappa(\lambda)$ number of partitioned subgraphs 
$$\mathcal{G}(\lambda) = \bigcup_{l=1}^{\kappa(\lambda)} 
G_l (\lambda) \; \mbox{   with } \; G_l =\{ V_l(\lambda), A_l(\lambda) \},$$
where $V_l$ and $A_l$ are node and edge sets of subgraph $G_l$.

Let $S= (s_{ij})$ be the sample covariance matrix. Let  $B(\lambda) = (b_{ij})$ be the adjacency matrix defined by
\bqn b_{ij}(\lambda) = 
\begin{cases}
1 &\; \mbox{  if } | \widehat{s}_{ij} | > \lambda;\\
0 & \; \mbox{ otherwise.}
\end{cases} \label{eq:Badj}\eqn

The adjacency matrix $B$ similarly induces a graph with $\tau(\lambda)$ disjoint subgraphs: 
$$\mathcal{H}(\lambda) = \bigcup_{l=1}^{\tau(\lambda)} H_l(\lambda) \; \mbox{ with }  H_l = \{W_l(\lambda), B_l(\lambda)\},$$
where $W_l$ and $B_l$ are node and edge sets of subgraph $H_l$.
Then the partitioned graphs are shown to be partially nested in a sense that the node sets exhibits persistency.

\begin{figure}[t]
\centering
\includegraphics[width=0.9\textwidth]{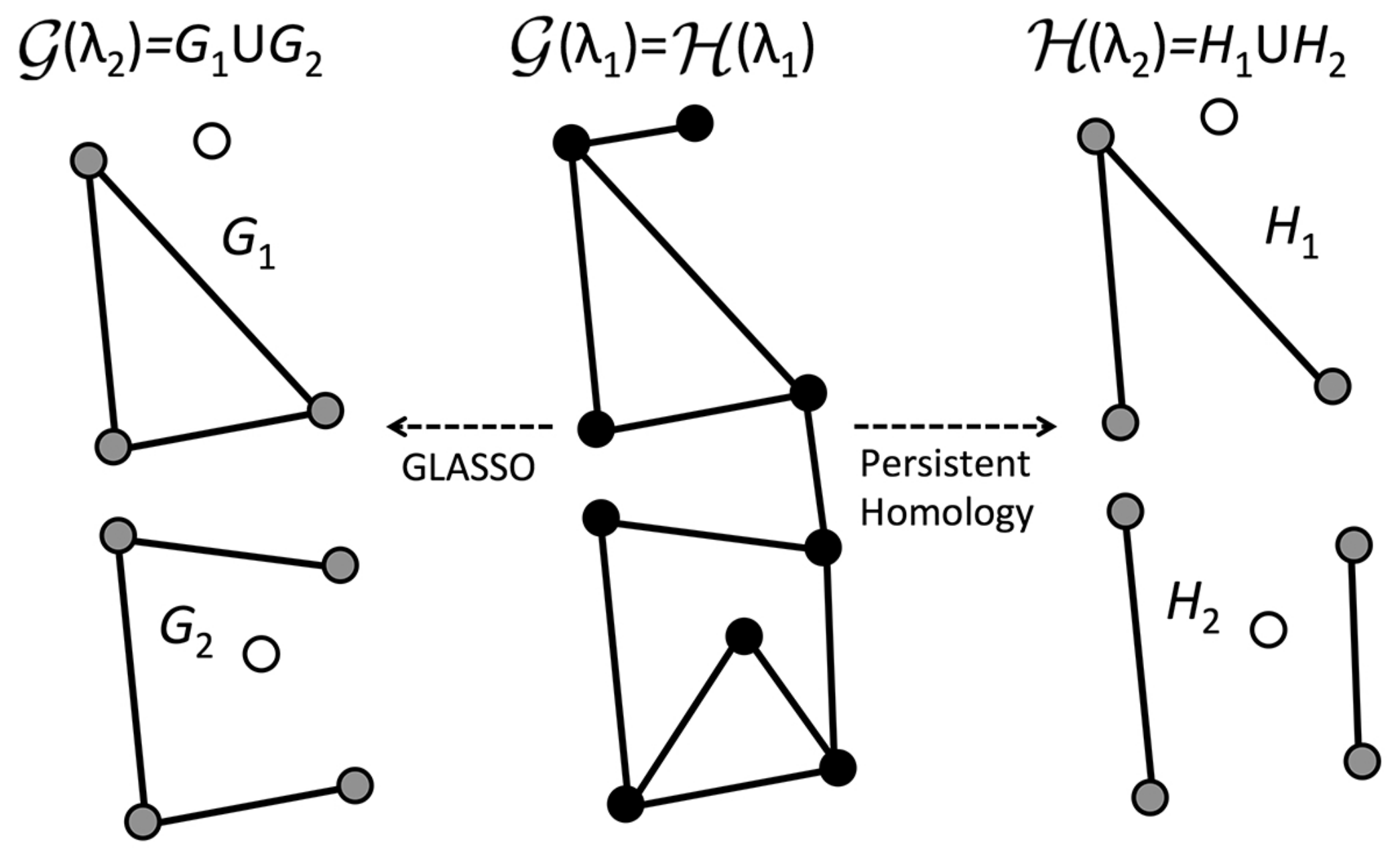}
 \caption{\footnotesize Schematic of graph filtrations obtained by sparse-likelihood (\ref{eq:Aadj}) and sample covariance thresholding (\ref{eq:Badj}). The vertex set of $\mathcal{G}(\lambda_1) = \mathcal{H}(\lambda_1)$ consists of black nodes. For the next filtration value $\lambda_2$, 
$\mathcal{G}(\lambda_2) \neq \mathcal{H}(\lambda_2)$ since the edge sets are different. However, the partitioned vertex sets (gray colored) of $\mathcal{G}(\lambda_2)$ and 
$\mathcal{H}(\lambda_2)$ match.}
\label{fig:glassofiltration}
\end{figure}

\begin{theorem}
\label{theorem:SL}
For any $\lambda >0$, the adjacency matrices (\ref{eq:Aadj}) and (\ref{eq:Badj}) induce the identical vertex partition so that $\kappa(\lambda)= \tau(\lambda)$ and $ V_l (\lambda)= W_l(\lambda)$. Further, the node sets $V_l$ and $W_l$ form filtrations over the sparse parameter:
\bqn V_l(\lambda_1) \supset V_l(\lambda_2) \supset V_l(\lambda_3)  \supset \cdots \label{eq:Vnested}\\
W_l(\lambda_1) \supset W_l(\lambda_2) \supset W_l(\lambda_3)  \supset \cdots\label{eq:Vnested2} \eqn
for $\lambda_1 \leq \lambda_2 \leq \lambda_3 \leq \cdots.$ 
\end{theorem}

From (\ref{eq:Badj}), it is trivial to see the filtration holds for $W_l$. The filtration for $V_l$ is proved in \citet{huang.2010}. The equivalence of the node sets $ V_l = W_l$ is proved in \citet{mazumder.2012}. Note that the edge sets may not form a filtration (Figure \ref{fig:glassofiltration}). The construction of the filtration on the node sets $V_l$ (\ref{eq:Vnested}) is very time consuming since we have to solve the sequence of graphical-LASSO. For instance, for 548 node sets and 547 different filtration values, the whole filtration takes more than  54 hours in a desktop \citep{chung.2015.TMI}.

\begin{figure}[t]
\includegraphics[width=1\linewidth]{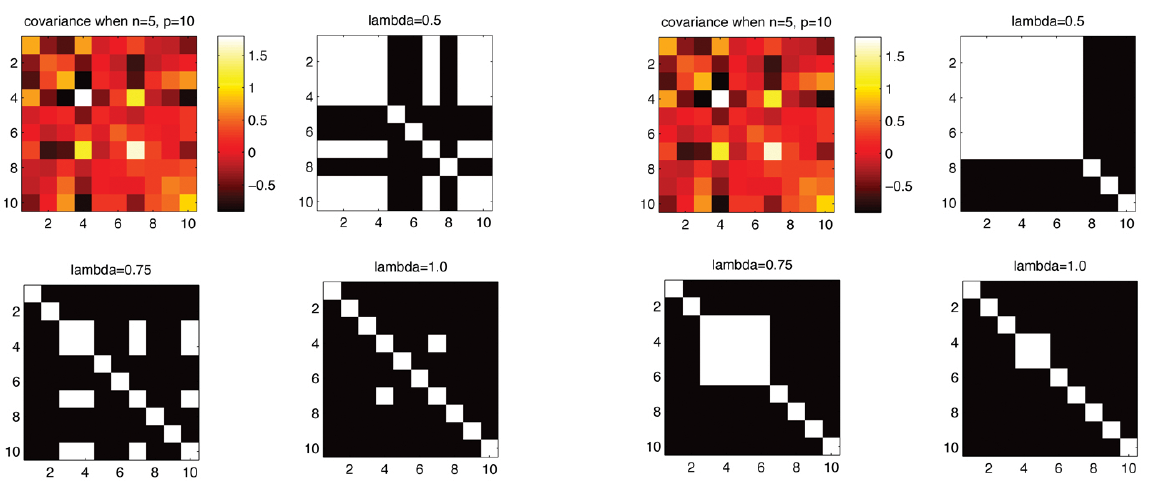}
\caption{Left: Adjacency matrices obtained through graphical-LASSO with increasing $\lambda$ values. The persistent homological structure is self-evident.  
Right: Adjacency matrices are clustered as a block diagonal matrix $D$ by permutation.} 
\label{fig:sparse-GLASSO}
\end{figure}

In Figure \ref{fig:sparse-GLASSO}, we randomly simulated the data matrix $X_{5 \times 10}$ from the standard normal distribution. The sample covariance matrix is then feed into graphical-LASSO with different filtration values. To identify the structure better, we transformed the adjacency matrix $A$ by permutation $P$ such that $D= P A P^{-1}$ is a block diagonal matrix. Theoretically only the partitioned node sets are expected to exhibit the nestedness but in this example, the  edge sets are also nested as well.

\section{Sparse correlation network}
\index{sparse correlation network}
\index{correlations!sparse}

The problem with graphical-LASSO or any type of similar L1 norm optimization is that it becomes computationally expensive as the number of node $p$ increases. So it is not really practical for large-scale brain networks. In this section, we present a scalable large-scale network model ($p >$ 25000) that yields greater computational speed and efficiency by bypassing the computational bottleneck of optimizing $L1$-penalties. 

There are few previous studies at speeding up the computation for sparse models. By identifying block diagonal structures in the estimated (inverse) covariance matrix, it is possible to reduce the computational burden in the penalized log-likelihood method \citep{mazumder.2012,witten.2011}.  
However, the method presented in this section differs from \citet{mazumder.2012} and \citet{witten.2011} in that we do not need to assume that the data to follow Gaussianness. Subsequently, there is no need to specify the likelihood function.  Further, the cost functions we are optimizing are different. Specifically, we propose a novel sparse network model based on {\em correlations}. Although correlations are often used in sciences in connection to times series and stochastic processes \citep{worsley.2005.neural,worsley.2005.royal}, the sparse version of correlation has been somewhat neglected.

Consider measurement vector ${\bf x}_j$ on node $j$. If we center and rescale the measurement ${\bf x}_j$ such that 
$$\parallel {\bf x}_j \parallel^2 = {\bf x}_{j}^{\top}{\bf x}_{j} =1,$$ 
the sample correlation between nodes $i$ and $j$ is given by
${\bf x}_{i}^{\top} {\bf x}_{j}$. Since the data is normalized, the sample covariance matrix is reduced to the sample correlation matrix.

Consider the following linear regression between nodes $j$ and $k$ $(k \neq j)$:
\bqn {\bf x}_j= \gamma_{jk} {\bf x}_k + \epsilon_j. \label{eq:LRG}\eqn
We are basically correlating data at node $j$ to data at node $k$. In this particular case, $\gamma_{jk}$ is the usual Pearson correlation. The least squares estimation (LSE) of $\gamma_{jk}$ is then given by 
\bqn \widehat{\gamma}_{jk} = {\bf x}_j^{\top}{\bf x}_k, \label{eq:gamma}\eqn
which is the sample correlation. For the normalized data, regression coefficient estimation is exactly the sample correlation. For the normalized and centered data, the regression coefficient is the correlation. It can be shown that (\ref{eq:gamma}) minimizes the sum of least squares over all nodes:
\bqn \sum_{j=1}^p \sum_{k \neq j}
\parallel {\bf x}_{j}  - \gamma_{jk} {\bf x}_{k} \parallel^{2}. \label{eq:LRG2}\eqn
Note that we do not really care about correlating ${\bf x}_j$ to itself since the correlation is then trivially $\gamma_{jj}=1$. \\

\subsection{Sparse correlations}

Let ${\bf \Gamma} = (\gamma_{jk})$ be the correlation matrix. The sparse penalized version of (\ref{eq:LRG2}) is given by
\bqn 
\label{eq:lasso_corr}
F ({\bf \Gamma})= 
\frac{1}{2}\sum_{j=1}^p \sum_{k \neq j}
\parallel {\bf x}_{j}  - \gamma_{jk} {\bf x}_{k} \parallel^{2} + \lambda \sum_{j=1}^p \sum_{k \neq j}  | \gamma_{jk} |.
\eqn
The sparse correlation is given by minimizing $F ({\bf \Gamma})$. By increasing $\lambda$, the estimated correlation matrix $\widehat{\bf \Gamma}(\lambda)$ becomes more sparse. When $\lambda=0$, the sparse correlation is simply given by the sample correlation, i.e. $\widehat{\gamma}_{jk} = {\bf x}_j^{\top}{\bf x}_k$.  As $\lambda$ increases, the correlation matrix ${\bf \Gamma}$ shrinks to zero and becomes more sparse. 
This is separable compressed sensing or LASSO  type problem. However, there is no need to numerically optimize (\ref{eq:lasso_corr})  using the coordinate descent learning or the active-set algorithm often used in compressed sensing  \citep{peng.2009,friedman.2008}. The minimization of (\ref{eq:lasso_corr}) can be done by the proposed soft-thresholding method analytically by exploiting the topological structure of the problem. Since ${\bf x}_{i}^{\top}{\bf x}_{j} \neq \delta_{ij}$, the Dirac delta, it looks like the sparse regression is not orthogonal design and the existing soft-thresholding method for LASSO \citep{tibshirani.1996} is not directly applicable. However, it can be made into orthogonal design. The detail is given in the sparse cross-correlation section.

\begin{theorem} 
\label{thm:softtresholding}
For $\lambda \geq 0$, the solution of the following separable LASSO problem
$$\widehat{\gamma}_{jk}(\lambda) = \arg \min_{\gamma_{jk}}  \frac{1}{2}\sum_{j=1}^p  \sum_{k \neq j} \parallel {\bf x}_{j}  - \gamma_{jk} {\bf x}_{k} \parallel^{2} + \lambda \sum_{j=1}^p  \sum_{k \neq j} | \gamma_{jk} |,$$
is given by the soft-thresholding
 \bqn \widehat{\gamma}_{jk}(\lambda) 
 = \begin{cases} 
 {\bf x}_{j}^{\top} {\bf x}_{k}  - \lambda & \mbox{ if }   {\bf x}_{j}^{\top} {\bf x}_{k}  > \lambda \\
 0                                               & \mbox{ if }   |{\bf x}_{j}^{\top} {\bf x}_{k}|  \le \lambda \\ 
{\bf x}_{j}^{\top} {\bf x}_{k}  + \lambda & \mbox{ if }   {\bf x}_{j}^{\top} {\bf x}_{k}  < -\lambda
\end{cases}.
\label{eq:lambda-without2}
\eqn
\end{theorem}

{\em Proof.} Write (\ref{eq:lasso_corr}) as
\bqn 
\label{eq:lasso_corr2}
F ({\bf \Gamma})= \frac{1}{2}\sum_{j=1}^p  \sum_{k \neq j} f(\gamma_{jk}),\eqn 
where 
$$f(\gamma_{jk}) = 
\parallel {\bf x}_{j}  - \gamma_{jk} {\bf x}_{k} \parallel^{2} + 2 \lambda  | \gamma_{jk} |.$$
Since $f(\gamma_{jk})$ is nonnegative and convex, $F ({\bf \Gamma})$ is minimum if each component $f(\gamma_{jk})$ achieves minimum. So we only need to minimize each component $f(\gamma_{jk})$.
This differentiates 
our sparse correlation formulation from the standard compressed sensing that cannot be optimized in this component wise fashion. $f(\gamma_{jk})$ can be rewritten as
\bq f(\gamma_{jk}) &=& 
\| {\bf x}_{j}\|^2   - 2 \gamma_{jk} {\bf x}_{j}^{\top} {\bf x}_{k} + \gamma_{jk}^2 \| {\bf x}_{k} \|^{2} + 2 \lambda  | \gamma_{jk} | \\
&=&(\gamma_{jk} - {\bf x}_{j}^{\top} {\bf x}_{k} )^2 + 2 \lambda  | \gamma_{jk} | +1.
\eq
We used the fact ${\bf x}_{j}^{\top} {\bf x}_{j} =1.$ 
  
For $\lambda=0$, the minimum of $f(\gamma_{jk})$ is achieved when
$\gamma_{jk} = {\bf x}_j^{\top} {\bf x}_k$, which is the usual LSE.
For $\lambda > 0$, 
Since $f(\gamma_{jk})$ is quadratic in $\gamma_{jk}$, the minimum is achieved when
\bqn 
\frac{\partial f}{\partial \gamma_{jk}} =  
 2 \gamma_{jk} -  2 {\bf x}_{j}^{\top} {\bf x}_{k}  \pm 2 \lambda =0 \label{eq:pm}\eqn
The sign of $\lambda$ depends on the sign of $\gamma_{jk}$. Thus, sparse correlation $\widehat{\gamma}_{jk}$ is given by a soft-thresholding of $ {\bf x}_{j}^{\top} {\bf x}_{k}$:
 \bqn \widehat{\gamma}_{jk}(\lambda) 
 = \begin{cases} 
 {\bf x}_{j}^{\top} {\bf x}_{k}  - \lambda & \mbox{ if }   {\bf x}_{j}^{\top} {\bf x}_{k}  > \lambda \\
0                                               & \mbox{ if }   |{\bf x}_{j}^{\top} {\bf x}_{k}|  \le \lambda \\ 
{\bf x}_{j}^{\top} {\bf x}_{k}  + \lambda & \mbox{ if }   {\bf x}_{j}^{\top} {\bf x}_{k}  < -\lambda
\end{cases}.
\label{eq:lambda-without2}
\eqn
$\square$

The  estimated sparse correlation (\ref{eq:lambda-without2}) basically thresholds the sample correlation that is larger or smaller than  $\lambda$ by the amount $\lambda$. Due to this simple expression,  there is no need to optimize (\ref{eq:lasso_corr}) numerically as often done in compressed sensing or LASSO \citep{peng.2009, friedman.2008}. However, Theorem \ref{thm:softtresholding} is only  applicable to separable cases and for non-separable cases, numerical optimization is still needed.

The different choices of sparsity parameter $\lambda$ will produce different solutions in sparse model $\mathcal{A}(\lambda)$. Instead of analyzing each model separately,  we can analyze  the whole collection of  all the sparse solutions for many different values of  $\lambda$. This avoids the problem of identifying the optimal sparse parameter that may not be optimal in practice. The question is then how to use the collection of $\mathcal{A}(\lambda)$ in a coherent mathematical fashion. This can be addressed using persistent homology  \citep{edelsbrunner.2008,lee.2011.MICCAI,lee.2012.TMI}.

\subsection{Filtration in sparse correlations}
\index{filtrations!sparse correlations}
\index{filtrations}

Using the sparse solution (\ref{eq:lambda-without2}), we can construct a filtration. We will basically build a graph $\mathcal{G}$ using spare correlations. Let $\widehat{\gamma}_{jk}(\lambda)$ be the sparse correlation estimate. Let $A(\lambda) = (a_{ij})$ be the adjacency matrix defined as
$$a_{jk}(\lambda) = 
\begin{cases}
1 &\; \mbox{  if  }  \widehat{\gamma}_{jk} (\lambda) \neq 0;\\
0 & \; \mbox{ otherwise.}
\end{cases}
$$
This is equivalent to the adjacency matrix $B=(b_{jk})$ defined as
\bqn b_{jk}(\lambda) = 
\begin{cases}
1 &\; \mbox{  if } |{\bf x}_j^{\top}{\bf x}_k | > \lambda;\\
0 & \; \mbox{ otherwise.}
\end{cases} \label{eq:Bcases}\eqn
The adjacency matrix $B$ is simply obtained by thresholding the sample correlations. Then the adjacency matrices $A$ and $B$ induce a identical graph $\mathcal{G}(\lambda)$ consisting of $\kappa(\lambda)$ number of partitioned subgraphs
$$\mathcal{G}(\lambda) = \bigcup_{l=1}^{\kappa(\lambda)} G_l (\lambda) \; \mbox{ with }  G_l = \{ V_l(\lambda), E_l(\lambda) \}, $$
where $V_l$ and $E_l$ are node and edge sets respectively. Note
$$\; G_l \bigcap G_m = \varnothing \; \mbox{ for any } \; l \neq m.$$
and no two nodes between the different partitions are connected.
The node and edge sets are denoted as $\mathcal{V}(\lambda) = \bigcup_{l=1}^{\kappa} V_l$ and $\mathcal{E}(\lambda) = \bigcup_{l=1}^{\kappa} E_l$ respectively. Then we have the following theorem: 

\begin{figure}[t]
\centering
\includegraphics[width=1\linewidth]{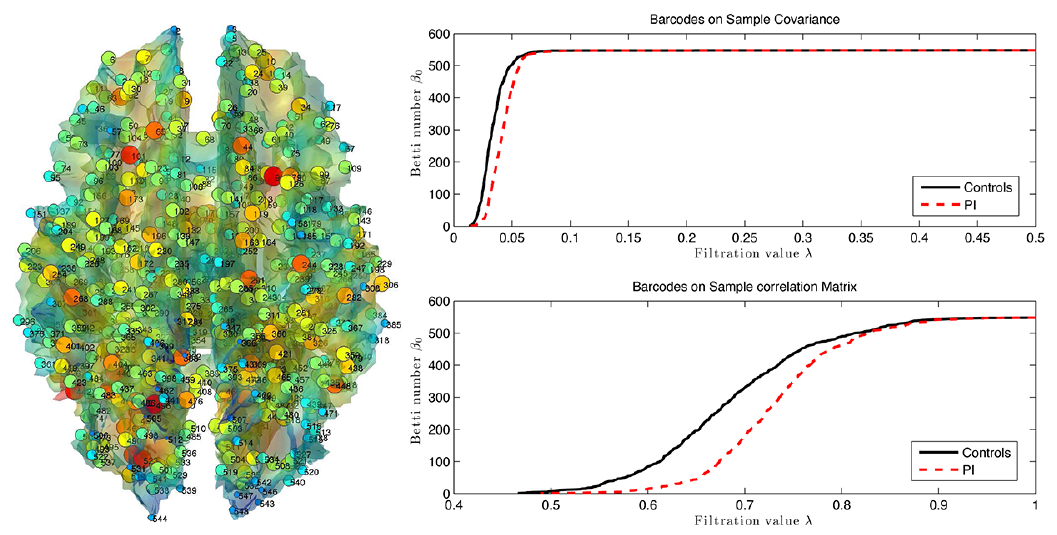}
\caption{Jocobian determinant of deformation field are measured at 548 nodes along the white matter boundary \citep{chung.2015.TMI}. The $\beta_0$-number (number of connected components) of the filtrations on the sample correlations and covariances show huge group separation between normal controls and post-institutionalized (PI) children. The topological patterns are similar regardless of the number of nodes used.} 
\label{fig:sparse-jacobian}
\end{figure}

\begin{theorem} The induced graph from the spare correlation forms a filtration:
\bqn \mathcal{G}(\lambda_1) \supset \mathcal{G}(\lambda_2) \supset \mathcal{G}(\lambda_3)  \supset \cdots 
\label{eq:PHG}
\eqn
for $\lambda_1 \leq \lambda_2 \leq \lambda_3 \leq \cdots$. Equivalently, the node and edge sets also form filtrations as well:
\bq  && \mathcal{V}(\lambda_1) \supset  \mathcal{V}(\lambda_2) \supset  \mathcal{V}(\lambda_3)  \supset \cdots \\
&& \mathcal{E}(\lambda_1) \supset  \mathcal{E}(\lambda_2) \supset  \mathcal{E}(\lambda_3). 
\label{eq:PHG2}
\eq
\end{theorem}
The proof can be easily obtained from the definition of adjacency matrix (\ref{eq:Bcases}).

\subsection{Sparse cross-correlations}
We can extend the sparse correlation framework to the sparse cross-correlations. Let $V= \{ v_1, \cdots, v_p \}$ be a node set where data is observed. We expect the number of nodes $p$ to be significantly larger than the number of images $n$, i.e., $p \gg n$. Let $x_k(v_i)$ and $y_k(v_i)$ be the $k$-th paired scalar measurements at node $v_i$. They can be twins, longitudinal scans or even multimodal images. Denote ${\bf x}(v_i) =(x_1(v_i), \cdots, x_n(v_i))^{\top}$ and ${\bf y}(v_i) =(y_1(v_i), \cdots, y_n(v_i))^{\top}$ be the paired data vectors over $n$ different images at voxel $v_i$.  Center and scale ${\bf x}$ and ${\bf y}$ such that
$$ \sum_{k=1}^n x_k(v_i) = \sum_{k=1}^n y_k (v_i) = 0, $$
$$\| {\bf x}(v_i) \|^2 = {\bf x}^{\top}(v_i){\bf x}(v_i) =  \| {\bf y}(v_i) \|^2  = {\bf y}^{\top}(v_i){\bf y}(v_i) = 1$$
for all $v_i$. 
The reasons for centering and scaling will soon be obvious.

\begin{figure}[t]
\centering
\includegraphics[width=0.7\linewidth]{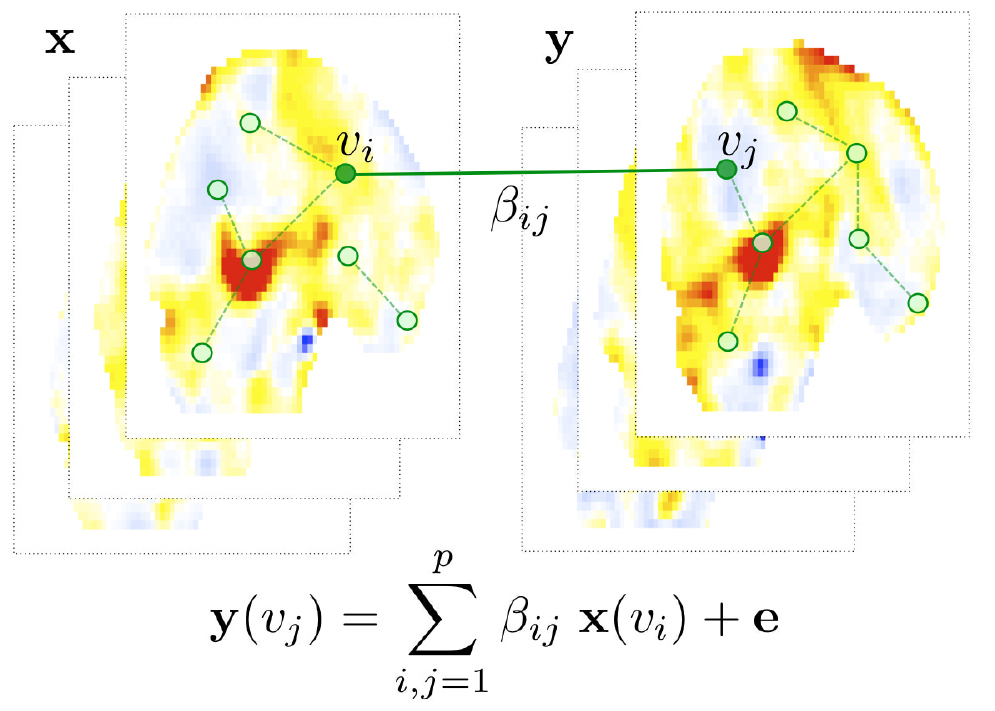}
 \caption{The schematic of hyper-network construction on paired image vectors ${\bf x}$ and ${\bf y}$. The  image vectors ${\bf y}$ at voxel $v_j$ is modeled as a linear combination of the first image vector ${\bf x}$ at all other voxels. The estimated parameters $\beta_{ij}$ give the hyper-edge weights.}\label{fig:hypernetwork}
\end{figure}

We set up a hyper-network by relating the paired vectors at different voxels $v_i$ and $v_j$:
\bqn {\bf y}(v_j)= \sum_{i=1}^p \beta_{ij} \; {\bf x}(v_i) + {\bf e} \label{eq:full}\eqn
for some zero-mean noise vector ${\bf e}$ (Fig. \ref{fig:hypernetwork}). The parameters $\beta=(\beta_{ij})$ are the weights of the hyper-edges between voxels $v_i$ and $v_j$ that have to be estimated. We are constructing a  physically nonexistent artificial network across different images. 
For fMRI, (\ref{eq:full}) requires estimating over billions of connections, which is computationally challenging.  
In practice however, each application will likely to force $\beta$ to have a specific structure that may reduce the computational burden. 

For this section, let us set up a linear model between ${\bf x}(v_i)$ and ${\bf y}(v_j)$:   
\bqn {\bf y}(v_j)= b_{ij} \; {\bf x}(v_i) + {\bf e}, \label{eq:LRG}\eqn
where ${\bf e}$ is the zero-mean error vector whose components are independent and identically distributed. 
Since the data are all centered, we do not have the intercept in linear regression (\ref{eq:LRG}). The least squares estimation (LSE) of $b_{ij}$ that minimizes the L2-norm 
\bqn \sum_{i,j=1}^p
\parallel {\bf y}(v_j) - b_{ij} \; {\bf x}({v_i}) \parallel^{2} \label{eq:LRG2}\eqn
is given by 
\bqn \widehat{b}_{ij} = {\bf x}^{\top}(v_i) {\bf y}(v_j),\label{eq:gamma}\eqn
which are the (sample)  {\em  cross-correlations} \citep{worsley.2005.neural,worsley.2005.royal}.  
 The cross-correlation is invariant under the centering and scaling operations. The sparse version of L2-norm (\ref{eq:LRG2}) is given by

\bqn
\label{eq:lasso_corr}
F(\beta; {\bf x}, {\bf y},\lambda) = \frac{1}{2} \sum_{i,j =1}^p 
\parallel {\bf y}(v_j) - \beta_{ij} \; {\bf x}({v_i}) \parallel^{2} + \lambda \sum_{i,j =1}^p  | \beta_{ij} |.
\eqn
The {\em sparse cross-correlation} is then obtained by minimizing over every possible $\beta_{ij} \in \mathbb{R}$: 
\bqn \widehat \beta (\lambda) = \arg \min_{\beta}  F(\beta; {\bf x}, {\bf y}, \lambda). \label{eq:betamin} \eqn
The estimated sparse cross-correlations $\widehat{\beta}(\lambda) = (\widehat \beta_{ij}(\lambda))$  shrink toward zero as sparse parameter $\lambda  \geq 0$ increases.  
The direct optimization of (\ref{eq:lasso_corr}) for large $p$ is computationally demanding. However, there is no need to optimize (\ref{eq:lasso_corr}) numerically using the coordinate descent learning or the active-set algorithm as often done in sparse optimization \citep{peng.2009,friedman.2008}. We can show that the minimization of (\ref{eq:lasso_corr}) is simply done algebraically.

\begin{theorem}
\label{theorem:SCC} 
For $\lambda \geq 0$, the minimizer of $F(\beta; {\bf x}, {\bf y}, \lambda)$ is given by 
 \bqn \widehat{\beta}_{ij}(\lambda) 
 = \begin{cases} 
 {\bf x}^{\top}(v_i) {\bf y}(v_j)  - \lambda & \mbox{ if }   \;  {\bf x}^{\top}(v_i) {\bf y}(v_j)  > \lambda \\
 0                                               & \mbox{ if }   \; | {\bf x}^{\top}(v_i) {\bf y}(v_j)|  \le \lambda \\ 
 {\bf x}^{\top}(v_i) {\bf y}(v_j)  + \lambda & \mbox{ if }   \;  {\bf x}^{\top}(v_i) {\bf y}(v_j) < -\lambda
\end{cases}.
\label{eq:cases}
\eqn
\end{theorem}

Although it is not obvious, Theorem \ref{theorem:SCC} is related to the orthogonal design in LASSO  \citep{tibshirani.1996} and the soft-shrinkage in wavelets \citep{donoho.1995}. To see this,  let us transform linear equations (\ref{eq:LRG}) into a index-free matrix equation:
\bq
\setlength{\arraycolsep}{0.2em}
\left[
\begin{array}{ccc}
 {\bf y}(v_1)   & \cdots & {\bf y}(v_1)  \\
{\bf y}(v_2)  & \cdots & {\bf y}(v_2)   \\
  \vdots &     \ddots & \vdots\\  
  {\bf y}(v_p)   & \cdots & {\bf y}(v_p)  
\end{array}
\right] 
= \left[
\begin{array}{cccc}
b_{11}{\bf x}(v_1)  &  b_{21}{\bf x}(v_2) & \cdots & b_{p1}{\bf x}(v_p)  \\
b_{12}{\bf x}(v_1)  &  b_{22}{\bf x}(v_2) & \cdots & b_{p2}{\bf x}(v_p)  \\
  \vdots &  \vdots &   \ddots & \vdots\\
b_{1p}{\bf x}(v_1)  &  b_{2p}{\bf x}(v_2) & \cdots & b_{pp}{\bf x}(v_p)  \\  
\end{array}
\right]  + \left[\begin{array}{ccc}
{\bf e}  &  \cdots & {\bf e}  \\
{\bf e}  &  \cdots & {\bf e}  \\
\vdots & \ddots  &  \vdots \\
{\bf e}  &  \cdots & {\bf e}  
\end{array}
\right].
\eq
The above matrix equation can be vectorized as follows.
\bq
\setlength{\arraycolsep}{0.1em}
\left[
\begin{array}{c}
{\bf y}(v_1)\\
  \vdots\\
{\bf y}(v_p)\\
\hline
\vdots\\
\hline
{\bf y}(v_1)\\
  \vdots\\
{\bf y}(v_p)
\end{array}
\right]
= \left[\begin{array}{ccc}
\begin{array}{ccc}
{\bf x}(v_1) & \cdots & 0 \\
\vdots & \ddots & \vdots \\
0 & \cdots & {\bf x}(v_1)
\end{array}
& \cdots &  \mbox{{\huge 0}}  \\
  \vdots & \ddots & \vdots\\
 \mbox{{\huge0}} & \cdots
 &
\begin{array}{ccc}
{\bf x}(v_p) & \cdots & 0 \\
\vdots & \ddots & \vdots \\
0 & \cdots & {\bf x}(v_p)
\end{array}
\end{array}\right]
\left[
\begin{array}{c}
b_{11}\\
\vdots\\
b_{p1}\\
\hline
\vdots\\
\hline
b_{1p}\\
\vdots\\
b_{pp}
\end{array}
\right]
 + 
\left[
\begin{array}{c}
{\bf e}\\
\vdots\\
{\bf e}\\
\hline
\vdots\\
\hline
{\bf e}\\
\vdots\\
{\bf e}
\end{array}
\right].
\eq

\begin{figure}[t]
\centering
\includegraphics[width=1\linewidth]{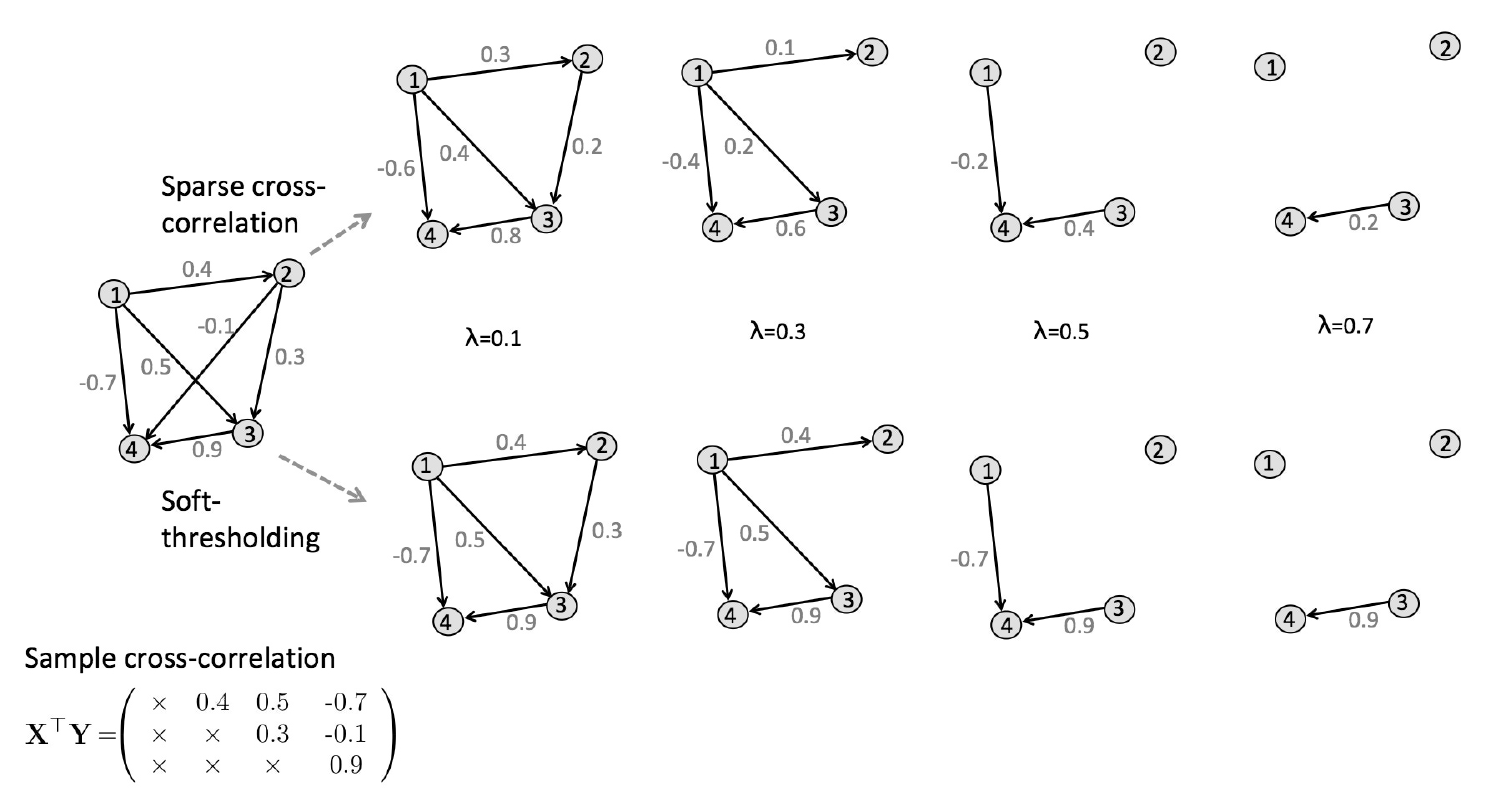} 
\caption{Schematic showing the equivalence of binary graph construction using the sparse cross-correlations and soft-thresholding. Top: The sparse cross-correlations are estimated  by minimizing the $L_1$ cost function (\ref{eq:lasso_corr}) for 4 different sparse parameters $\lambda$. The edge weights shrinked to zero are removed. Bottom: the equivalent binary graph can be obtained by soft-thresholding, i.e., simply thresholding the sample cross-correlations at $\lambda$.} 
\label{fig:HI-illustration}
\end{figure}
\begin{figure}[t]
\centering
\includegraphics[width=0.7\linewidth]{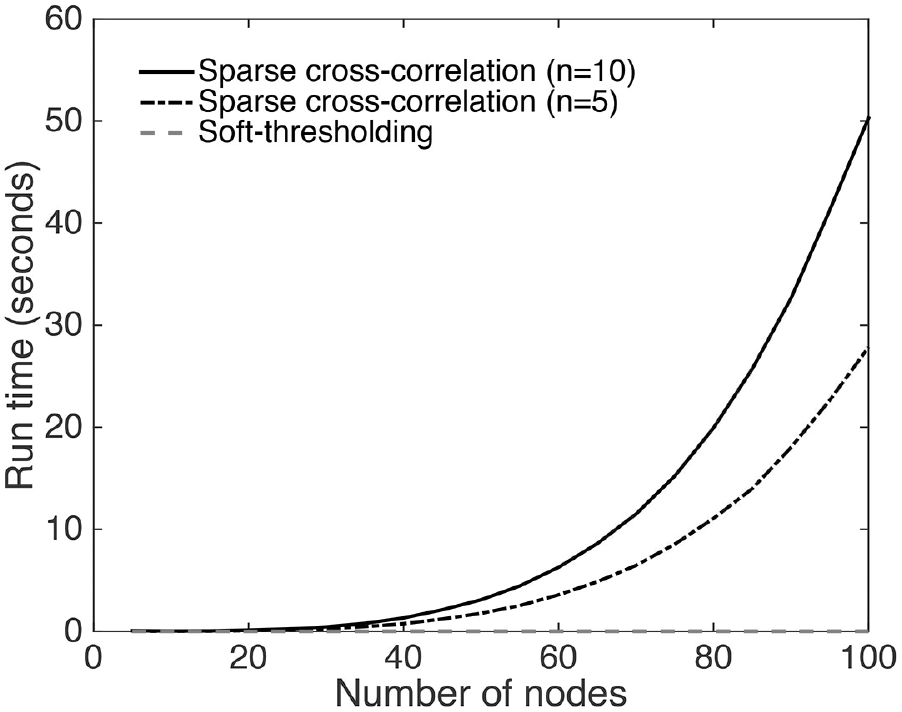}
\caption{Run time comparison of estimating sparse cross-correlations. The LASSO-based numerical optimization  with $n=5,10$ images with varying number of nodes. The run time scale linearly with the number of images but scale exponentially with the number of nodes. The LASSO runs more than 100000 times slower compared to the soft-thresholding method for $p=100$ nodes and $n=10$ images.} 
\label{fig:sparse-runtime}
\end{figure}

The above equation can be written in a more compact form. Let
\bq {\bf X}_{n \times p} &=& [{\bf x}(v_1) \; {\bf x}(v_2) \cdots \; {\bf x}(v_p) ]\\
{\bf Y}_{n \times p} &=& [{\bf y}(v_1) \; {\bf y}(v_2) \cdots \; {\bf y}(v_p) ]\\
{\bf 1}_{a \times b} & = & 
\left[\begin{array}{cccc}
1 &1 &  \cdots & 1\\
\vdots & \vdots & \ddots & \vdots\\
1 & 1 & \cdots & 1
\end{array}\right]_{a \times b}.
\eq
Then the matrix equation can be written as 
\bqn {\bf 1}_{p \times 1} \otimes vec({\bf Y}) =  \mathbb{X}_{np^2 \times p^2}\; vec(b) + {\bf 1}_{np^2 \times 1} \otimes {\bf e}, \label{eq:gigantic}\eqn
 where {\em vec} is the vectorization operation. The block diagonal design matrix $\mathbb{X}$ consists of $p$ diagonal blocks $I_p \otimes {\bf x}(v_1), \cdots, I_p \otimes {\bf x}(v_p)$, where $I_{p}$ is $p \times p$ identity matrix. Subseqeuntly, 
 $\mathbb{X}^{\top}\mathbb{X}$ is again a block diagonal matrix, where the $i$-th block is
 $$[I_p \otimes {\bf x}(v_i)]^{\top}[I_p \otimes {\bf x}(v_i)] = I_p \otimes [{\bf x}(v_i)^{\top} {\bf x}(v_i)] = I_p.$$
Thus, $\mathbb{X}$ is an orthogonal design. However, our formulation is {\em not} exactly the orthogonal design of LASSO as specified in  \citep{tibshirani.1996} since the noise components in (\ref{eq:gigantic}) are not independent. Further in standard LASSO, there are more columns than rows in  $\mathbb{X}$. In our case, there are $n$ times more rows. Still the soft-thresholding method introduced in  \citep{tibshirani.1996} is applicable and we obtain the analytic solution, which speed up the computation drastically compared to existing LASSO-based numerical optimization  (Figure \ref{fig:sparse-runtime}) \citep{peng.2009,friedman.2008}.
 
Theorem \ref{theorem:SCC} generalizes the sparse correlation case given in \citet{chung.2013.MICCAI}. 
Figure \ref{fig:HI-illustration}-top displays an example of obtaining sparse cross-correlations from the initial sample cross-correlation matrix
$${\bf X}^{\top}{\bf Y}=
\left(
\begin{array}{cccc}
 \times & \mbox{0.4}  & \mbox{0.5} & \mbox{ -0.7} \\
 \times&  \times &  \mbox{0.3} & \mbox{ -0.1} \\
   \times &   \times &  \times  & \mbox{ 0.9}
\end{array}
\right)
$$
using Theorem \ref{theorem:SCC}. Due to directional nature of the cross-correlation matrix, only the upper triangle part of the sample cross-correlation is demonstrated.

\section{Partial correlation network}
\index{networks!partial correlation}
\index{partial correlation}
\index{correlations!partial}

Let $p$ be the number of nodes in the network. In most applications, the number of nodes is expected to be larger than the number of observations $n$, which gives an underdetermined system.  Consider measurement vector at the $j$-th node 
$${\bf x}_j = (x_{1j}, \cdots, x_{nj})^{\top}$$
consisting of $n$ measurements. Vector ${\bf x}_j$ are assumed to be distributed with mean zero and covariance  
$\Sigma =(\sigma_{ij})$.  The correlation $\gamma_{ij}$ between the two nodes $i$ and $j$ is given by $$\gamma_{ij} = \frac{\sigma_{ij}}{\sqrt{\sigma_{ii} \sigma_{jj}}}.$$ By thresholding the correlation, we can establish  a link between two nodes. However, there  is a problem with this simplistic approach in that it fails to explicitly factor out the confounding effect of other nodes. To remedy   this problem, partial correlations can be used in factoring out the dependency of other nodes \citep{he.2007, marrelec.2006, huang.2009, huang.2010,peng.2009}.

If we denote the inverse covariance matrix as $\Sigma^{-1} = (\sigma^{ij})$, the {\em partial correlation} between the nodes $i$ and $j$ while factoring out the effect of all other nodes is given by \citep{peng.2009}
\bqn \rho_{ij} = -\frac{\sigma^{ij}}{\sqrt{\sigma^{ii} \sigma^{jj}}}. \label{eq:rho}\eqn
Equivalently, we can compute the partial correlation  {\em via} a linear model as follows. Consider a linear model of correlating measurement at node $j$ to all other nodes:
\bqn {\bf x}_{j} = \sum_{k \neq j}   \beta_{jk}  {\bf x}_{k} + \epsilon_k. \label{eq:xpartial} \eqn 
The parameters $\beta_{jk}$ are estimated by minimizing the sum of squared residual of (\ref{eq:xpartial}) 
\bqn L(\beta) = \sum_{j=1}^p \| {\bf x}_j   - \sum_{k \neq j} \beta_{jk} {\bf x}_k \|^2 \label{eq:Lbeta}\eqn
in a least squares fashion. If we denote the least squares estimator by $\widehat{\beta_{jk}}$, the residuals are given by  
\bqn {\bf r}_j = {\bf x}_{ j} - \sum_{k \neq j}   \widehat{\beta_{jk}}  {\bf x}_{k}.\label{eq:xresidual} \eqn
The partial correlation is then obtained by computing the correlation between the residuals  \citep{ he.2007, lerch.2006, peng.2009}:
$$\rho_{ij} = \mbox{corr } ({\bf r}_i, {\bf r}_j).$$

\subsection{Sparse partial correlations} 
There is a serious problem with the least squares estimation framework discussed in the previous section. Since $n \ll p$, this is a significantly underdetermined system. This is also related to 
 the covariance matrix $\Sigma$ being singular so we cannot just invert the covariance matrix. For this, we need sparse network modeling.
 
 The minimization of (\ref{eq:Lbeta}) is exactly given by solving the normal equation:
\bqn {\bf x}_j   = \sum_{k \neq j} \beta_{jk} {\bf x}_k, \label{eq:normal2}\eqn 
which can be  turned into standard linear form $y= A \beta$ \citep{lee.2011.TMI}. 
Note that (\ref{eq:normal2}) can be written as
$${\bf x}_j =\underbrace{[{\bf x}_1, \cdots, {\bf x}_{j-1}, {\bf 0}, {\bf x}_{j+1},  \cdots, {\bf x}_p]}_{{\bf X}_{-j}}
\underbrace{\left( 
\begin{array}{c} \beta_{j1}\\ \beta_{j2} \\ \vdots \\ \beta_{jp} \end{array} 
\right)}_{\beta_j},$$
where ${\bf 0}_{n \times 1}$ is a column vector of all zero entries. Then we have
\bqn
\underbrace{\left( \begin{array}{c} {\bf x}_1\\ {\bf x}_2 \\ \vdots \\ {\bf x}_p \end{array} \right)}_{y_{np \times 1}}
=
\underbrace {
\left(
\begin{array}{cccc}
 {\bf X}_{-1} & {\bf 0}  &  \cdots & {\bf 0} \\
 {\bf 0} &  {\bf X}_{-2} &  \cdots & {\bf 0} \\
\vdots & \vdots & \ddots & \vdots\\
{\bf 0}  & {\bf 0}  & \cdots & {\bf X}_{-p}   
\end{array}
\right)}_{A_{np \times p^2}}  
\underbrace{\left( \begin{array}{c} \beta_1\\ \beta_2 \\ \vdots \\ \beta_p \end{array} \right)}_{\beta_{p^2 \times 1}}, \label{eq:bigCS}\eqn
where $A$ is a block diagonal matrix and  ${\bf 0}_{n \times p}$ is a matrix of all zero entries. 
We regularize (\ref{eq:bigCS}) by incorporating $l_1$ LASSO-penalty $J$ \citep{tibshirani.1996, peng.2009, lee.2011.TMI}:
$$J = \sum_{i,j} | \beta_{ij} |.$$
The sparse estimation of $\beta_{ij}$ is then given by minimizing $L + \lambda J$. Since  there is dependency between $y$ and $A$,  (\ref{eq:bigCS}) is not exactly a standard compressed sensing problem \citep{peng.2009, lee.2011.TMI}. It should be intuitively understood that sparsity makes the linear equation (\ref{eq:normal2}) less underdetermined. The larger the value of $\lambda$, the more sparse the underlying topological structure gets. Since
$$\rho_{ij} = \beta_{ij} \sqrt{ \frac{\sigma^{ii}}{\sigma^{jj}}},$$
the sparsity of $\beta_{ij}$ directly corresponds to the sparsity of $\rho_{ij}$, which is the strength of the link between nodes $i$ and $j$ \citep{peng.2009, lee.2011.TMI}. Once the sparse partial correlation matrix $\rho$ is obtained, we can simply link nodes $j$ and $j$, if $\rho_{ij} > 0$ and assign the weight $\rho_{ij}$ to the edge. This way, we obtain the weighted graph.

\subsection{Limitations}

However, the sparse partial correlation framework has a serious computational bottleneck. For $n$ measurements over $p$ nodes, it is required  that we solve a linear system with  an extremely large $A$ matrix of size $np \times p^2$, so that the complexity of the problem increases by  a factor of $p^3$!   Consequently, for a large number of nodes, the problem immediately becomes almost intractable for a small computer. For  example, for 1 million nodes, we have to compute $1$ trillion possible pairwise relationships between nodes. One practical solution is to modify (\ref{eq:xpartial}) so that the measurement at node $i$ is represented more  sparsely over some possible index set $S_i$:
$$x_i = \sum_{S_i} \beta_{ij} x_j + \epsilon_i.$$
making the problem substantially smaller. 

An alternate approach is to simply follow the {\em homotopy path}, which adds network edges one by one with a very limited increase of computational complexity so there is no need to compute $\beta$ repeatedly from scratch \citep{donoho.2006, plumbley.2005, osborne.2000}. 
The trajectory of the optimal solution $\beta$ in LASSO  follows a piecewise linear path as we change $\lambda$. By tracing the linear path, we can substantially reduce the computational burden of reestimating $\beta$ when $\lambda$ changes.

\section*{Acknowledgements}
The part of this study was supported by NIH grants NIH R01 EB022856 and R01 EB028753. We would like to thank Anqi Qiu of National University of Singapore for providing the data used in Figure 1 and 2, and Seth Pollak of University of Wisconsin-Madison for the data used in Figure 5. 
\bibliographystyle{agsm} 
\bibliography{reference.2020.07.22}

\end{document}